**Current-driven magnetization switching in a van der Waals ferromagnet Fe₃GeTe₂**

Xiao Wang†[1,2], Jian Tang†[1,2], Xiuxin Xia†[3,4], Congli He[5], Junwei Zhang[6,7], Yizhou Liu[1,2], Caihua Wan[1,2], Chi Fang[1,2], Chenyang Guo[1,2], Wenlong Yang[1,2], Yao Guang[1,2], Xiaomin Zhang[1,2], Hongjun Xu[1,2,8], Jinwu Wei[1,2,8], Mengzhou Liao[1,2], Xiaobo Lu[1,2], Jiafeng Feng[1,2], Xiaoxi Li[3,4], Yong Peng[7], Hongxiang Wei[1,2], Rong Yang[1,2,8], Dongxia Shi[1,2,8], Xixiang Zhang[6], Zheng Han,[3,4]*, Zhidong Zhang[3,4], Guangyu Zhang[1,2,8]* Guoqiang Yu[1,2,8]* and Xiufeng Han[1,2,8]

[1]Beijing National Laboratory for Condensed Matter Physics, Institute of Physics, Chinese Academy of Sciences, Beijing 100190, China

[2]Center of Materials Science and Optoelectronics Engineering, University of Chinese Academy of Sciences, Beijing 100049, China

[3]Shenyang National Laboratory for Materials Science, Institute of Metal Research, Chinese Academy of Sciences, Shenyang, China

[4]School of Material Science and Engineering, University of Science and Technology of China, Anhui 230026, China

[5]Institute of Advanced Materials, Beijing Normal University, Beijing 100875, China

[6]Physical Science and Engineering Division (PSE), King Abdullah University of Science and Technology (KAUST), Thuwal 23955-6900, Saudi Arabia

[7]Key Laboratory for Magnetism and Magnetic Materials of Ministry of Education, Lanzhou University, Lanzhou 730000, People's Republic of China

[8]Songshan Lake Materials Laboratory, Dongguan, Guangdong 523808, China

†These authors contributed equally to this work.

Email address: vitto.han@gmail.com, gyzhang@iphy.ac.cn, guoqiangyu@iphy.ac.cn

**The recent discovery of ferromagnetism in two-dimensional (2D) van der Waals (vdW) materials holds promises for novel spintronic devices with exceptional performances. However, in order to utilize 2D vdW magnets for building spintronic nanodevices such as magnetic memories, key challenges remain in terms of effectively switching the magnetization from one state to the other electrically. Here, we devise a bilayer structure of $Fe_3GeTe_2$/Pt, in which the magnetization of few-layered $Fe_3GeTe_2$ can be effectively switched by the spin-orbit torques (SOTs) originated from the current flowing in the Pt layer. The effective magnetic fields corresponding to the SOTs are further quantitatively characterized using harmonic measurements. Our demonstration of the SOT-driven magnetization switching in a 2D vdW magnet could pave the way for implementing low-dimensional materials in the next-generation spintronic applications.**

# Introduction

Emerging phenomena arising from the interfaces and heterostructures of conventional magnetic thin films[1, 2], such as exchange bias[3], interfacial perpendicular magnetic anisotropy (PMA)[4, 5], spin-transfer torque[6-8] and spin-orbit torques (SOTs)[9, 10], have greatly advanced the development of spintronic applications[8, 11]. Yet the pursuit of new magnetic materials with better interfacial properties and thinner thicknesses is still one of the main themes in spintronic studies. Van der Waals (vdW) materials offer a versatile platform for exploring novel phenomena and can provide high-quality interfaces at the atomic scale. Their intersection with spintronics has just been properly established with the recently discovered 2D magnetism[12-21]. It is believed that spintronic devices harnessing vdW magnets may inherit many advantages of 2D materials such as the gate tunability, flexibility, low-cost/large-scale growth, etc. However, studies on manipulating the magnetic order parameter of vdW magnets via spintronic approaches, which is essential for practical applications, have been rarely studied so far.

A feasible scheme to spintronically control the magnetization of vdW magnets is to combine their PMA with the effect of SOTs, by employing an additional heavy metal layer next to the magnetic layer. In this work, we demonstrate a SOT-driven perpendicular magnetization switching in a bilayer structure of few-layered $Fe_3GeTe_2$ (FGT) and Pt (Figs. 1a and 1b). FGT is currently one of the most attractive 2D magnetic vdW materials due to its gate-tunable Curie temperature $T_c$ (up to room temperature) and PMA[20, 21], which are both important for high-density information storage applications[4]. The crossover between magnetic vdW materials and SOTs opens the possibilities to push spintronic devices to the 2D limit with faster speed and lower energy consumption.

## Results

We fabricated the FGT/Pt device by first exfoliating few-layered FGT flakes from a high-quality bulk crystal onto Si/SiO$_2$ substrates. Characterizations of the magnetic properties and the structure of bulk FGT are shown in supplementary Section S1. Figure 1c shows a representative high angle annular dark field scanning transmission electron microscopy (HAADF-STEM) image of the as-exfoliated FGT. The atomic arrangements in the [100] axes are in agreement with the crystal structure of FGT. Each layer is constituted by the alternately arranged Te−Fe−Ge(Fe)−Fe−Te atomic planes. The vdW gaps (dark area) are visible between different layers. The exfoliated FGT flakes show a minimum step height of 0.8 nm on the surface (Figs. 1d and e), which matches an atomic layer thickness of FGT. After exfoliating the FGT flakes, the substrates were immediately transferred into a high-vacuum sputtering system and a 6 nm-thick Pt layer was then deposited on top of FGT. The FGT/Pt bilayers were then patterned into Hall bar devices (see Fig. 1f) by the processes described in supplementary Section S2.

We first measured the resistance of the device as a function of temperature, as shown in Fig. 1g. The bilayer device exhibits a metallic behavior. To elucidate the resistance behavior of FGT, the resistance of Pt layer is separately characterized by measuring a Hall bar control device prepared on a Si/SiO$_2$ substrate. After removing the resistance contribution from the Pt layer (see Fig. S14), the resistance of FGT can be roughly extracted (see Fig. 1g). The temperature dependence of the FGT resistance is similar to the previous reports for FGT with 4 to 6 layers (3.2 - 4.8 nm) [20]. The actual layer number observed by TEM is larger, as shown in Fig. 1b. We attribute this difference to the oxidization of FGT.

Magnetic properties of our devices were then characterized by measuring the Hall resistances. Figure 2a shows the Hall resistance as a function of the out-of-plane magnetic fields at various temperatures. Due to the intrinsic magnetization of FGT, anomalous Hall resistance dominates over the total Hall resistance below $T_c$ as indicated by the square-shaped loops. The hysteresis loop gradually disappears and the Hall resistance curve becomes more linear with increasing the temperature. The $T_c$ of our device is determined to be ~158 K by performing the Arrott plots[20, 22], as shown in Fig. 2b. Comparing to the previously reported layer-dependent $T_c$[9], the obtained $T_c$ in our FGT corresponds to ~5 layers (4 nm). This thickness value is in consistent with the estimations from the resistance measurements and the TEM image. Below $T_c$, the device exhibits PMA, as manifested by the much larger saturation field in the in-plane direction than that in the out-of-plane direction (Fig. 2c).

Next, we show that a current flowing in the bilayer can generate SOTs, which are originated from the spin Hall effect in Pt and/or the interfacial effects. The SOTs are characterized through harmonic measurements[23, 24]. For the harmonic measurements, a small a.c. current is applied to the device in the presence of an in-plane external magnetic field along the longitudinal (transverse) direction for measuring longitudinal (transverse) effective fields. Figures 3a-b and Figures 3d-e show the measured harmonic voltages under longitudinal ($H_L$) and transverse ($H_T$) external magnetic fields, which are fitted by parabolic and linear functions, respectively. The ratios corresponding to damping-like and field-like torques can be calculated as $B_{L(T)} = -2\left(\frac{\partial V_{2\omega}}{\partial H_{L(T)}}\right) / \frac{\partial^2 V_\omega}{\partial H_{L(T)}^2}$. Here, $V_\omega$ and $V_{2\omega}$ are the first and second harmonic voltage, respectively. The current-induced effective fields in the longitudinal and transverse directions can then be extracted based on[25] $\Delta H_{L(T)} = (B_{L(T)} \pm 2\beta B_{T(L)})/(1 - 4\beta^2)$. Here, $\beta$ is the ratio between the planar Hall resistance ($R_P$) and the anomalous Hall resistance ($R_{AHE}$) and $\beta = R_P/R_{AHE} = 0.12$. The obtained effective fields versus current density are $\Delta H_L = 53.4 \pm 4.7$ mT per $10^7$ A/cm$^2$ and $\Delta H_T = 24.3 \pm 2.3$ mT per $10^7$ A/cm$^2$, respectively. Note that the thermal contribution has been carefully considered (see supplementary Section S3). The observed damping-like torque and field-like torque are much larger than their typical values in the Pt/transition ferromagnetic metal bilayer structures[23, 26].

In general, the SOT can be evaluated by the spin torque efficiency[27, 28]: $\xi = T_{int}\theta_{SH} = \frac{2e}{\hbar}\mu_0 M_s t_{FM}^{eff} \Delta H_{L(T)}/J_e$, where $T_{int}$ is the interfacial spin transparency ($T_{int} <$ 1), $\theta_{SH}$ is the spin Hall angle, $e$ is the electron charge, $\hbar$ is the reduced Planck constant, $\mu_0$ is the vacuum permeability, $M_s$ is the saturation magnetization, $t_{FM}^{eff}$ is the effective thickness of FGT, and $J_e$ is the current density. The bulk $M_s$ value of FGT is approximately estimated to be 321 MA/m (see Fig. S7). If using this $M_s$ value, the estimated values of $\xi$ are unusually large (2.48 for the damping-like part and 1.14 for the field-like part, respectively) in comparison with the previously reported $\xi$ values for Pt (~0.12 for the damping-like part at room temperature)[27, 28]. As $\xi$ is only determined by the spin Hall angle of Pt and the interface quality, its value is unlikely to be one order of magnitude greater than 0.12. This discrepancy can be explained by the drop of $M_s$ due to the finite size effect and strong surface modification in ultra-thin magnetic films[29, 30]. Furthermore, $\xi$ may shed a light on estimating the upper limit of $M_s$ for few-layered vdW magnets. If assume a maximum spin torque efficiency $\xi = 0.12$, we can approximately estimate the upper limit value of $M_s$, which is $M_s = 16$ MA/m. This value is significantly smaller than the bulk $M_s$ value (321 MA/m, see Fig. S7).

Finally, we show that the generated SOTs can be used to switch the perpendicularly magnetized FGT in the bilayer devices with the assistance of an in-plane magnetic field to break the mirror symmetry[31]. Figures 4a and b show the current-induced magnetization switching of FGT with the in-plane magnetic field $H_x = 50$ mT and $-50$ mT at 100 K, respectively. The magnetization of FGT can be switched from one state to the other by sweeping the electric current. Two distinct states can be well sustained at the zero current. The switching polarity is anticlockwise (clockwise) for the positive (negative) in-plane field, indicating a positive spin Hall angle of Pt here, which is consistent with the previous works. Similar behaviors can be observed from 10 K to 130 K (see supplementary section S4), suggesting a large temperature window for switching. The switching behavior disappears around 140 K (still below $T_c$), which is likely due to the gradual loss of squareness of the hysteresis loops and the resulting absence of two distinct remanence states.

## Discussion

It is noted that the two observed resistance states during the switching are not fully saturated (indicated by the dashed lines in Figs. 4**a** and **b**). This incomplete switching can be ascribed to a Joule heating effect. As shown in Fig. 4**c**, starting from a fully saturated initial state (polarized in the positive direction), the magnetization of FGT (represented by the Hall resistance $R_{xy}$) begins to decrease once the applied current exceeds 8 mA. In our device, the positive current favors a positive magnetization under a positive in-plane magnetic field, so $R_{xy}$ should not decrease with increasing the current. Moreover, the decreasing behavior is independent on the initial state (supplementary Section S5), indicating its thermal origin. We have further extracted the temperature of the device under applied current pulses (with 50 ms duration) by monitoring the resistance of the device (blue squares in Fig. 4c)[32]. A significant thermal effect is observed, as the device temperature is approaching $T_c$ under a current of 8 mA and exceeding $T_c$ under a current of 10.5 mA. When the device temperature is close to $T_c$, the magnetic interactions of FGT are not sufficient to against the thermal fluctuations and maintain a single domain state. Thus, multi-domain state is formed under large applied current pulses, which results in the unsaturated resistance sates. Further increasing the current to above 10.5 mA also does little to help realizing a complete switching since the temperature rises above $T_c$.

To further prove the switching is originated from the current-induced SOT, we have also fabricated FGT/Ta bilayer devices and observed similar switching behavior but with opposite switching polarity due to the opposite sign of the spin Hall angle in Ta, as shown in Fig. S30. Figure 4d shows the switching diagram for various temperatures. The switching current decreases with the increase of temperature. We attribute the reduction of switching current to the simultaneous decrease in $M_s$ (manifested by the decrease of $R_{AHE}$) and effective PMA field ($H_k$) (see supplementary Section S6).

In summary, a SOT-driven perpendicular magnetization switching has been demonstrated in a FGT/Pt bilayer device. Our proof-of-concept SOT devices highlight the potential of magnetic vdW materials and their compatibilities with spintronic technologies. Further work is still needed to push the FGT down to the monolayer limit. Other than the Pt or Ta used in this work, vdW materials with strong spin-orbit coupling and non-trivial electronic properties can also be employed as the SOT sources, leading to the possible all-vdW magnetic memories. The large family of vdW materials and numerous combinations of vdW heterostructures remarkably extend the material choices and can be visioned as new building blocks for spintronic applications in the near future.

## Materials and Methods

### Growth and characterization of FGT bulk crystal

Single crystals of FGT were prepared by chemical vapor transport (CVT) method with iodine as the transport agent. High-purity (99.99%) Fe, Ge, and Te were milled into powder form with a stoichiometric molar proportion of 3:1:2 (Fe: Ge: Te) in an agate mortar.

### Device fabrication

We obtain few-layered FGT flakes with a freshly-cleaved surface on Si/SiO$_2$ substrate by our new developed scratching method (refer to the supplementary Section S2) through our home-made transfer station. Next, we deposited a 6 nm-thick Pt (or Ta(6 nm)/Pt(1.5 nm)) layer on FGT surface and the obtained FGT/Pt bilayer is air-stable and this could confirm the subsequent fabrication process. The Pt/FGT bilayer was patterned as Hall geometry by a standard e-beam lithography process and excrescent areas were etched by ion milling. Finally, the electrodes were fabricated by Ti(3 nm)/Au(50 nm) through e-beam evaporation.

### Characterizations

We used aberration-corrected scanning transmission electron microscopy (AC-STEM) to image the cross-sections directly. The cross-sectional samples were fabricated by focused ion beam (FIB) cutting along the [100] axes of Fe$_3$GeTe$_2$. All the electrical measurements were performed in a Physical Property Measurement System (PPMS) system with magnetic fields of up to 9 T and temperatures down to 1.8 K. Multiple lock-in-amplifiers (Stanford SR830 and SR850) and Keithley source meters (Keithley 2400, 2182, and 6221) were connected to the PPMS, enabling comprehensive transport measurements for the Hall bar devices. A constant of 200 $\mu$A d.c. current was applied for the Hall measurement.

**Acknowledgments**

G.Y. and X.H. thank the finical supports from the National Key Research and Development Program of China (Grant No. 2017YFA0206200, 2018YFB0407600, 2016YFA0300802, 2017YFA0206302), the National Natural Science Foundation of China (NSFC, Grants No.11874409, 11804380, 11434014, 51831012), the NSFC-Science Foundation Ireland (SFI) Partnership Programme (Grant No. 51861135104), and 1000 Youth Talents Program. G.Z. thanks the finical supports from NSFC (Grant Nos. 61734001, 11834017 and 51572289), the Strategic Priority Research Program (B) of CAS (Grant No. XDB30000000), the Key Research Program of Frontier Sciences of CAS (Grant No. QYZDB-SSW-SLH004), the National Key R&D program of China (Grant No. 2016YFA0300904). Y.L. acknowledges support from the Institute of Physics, Chinese Academy of Sciences through the International Young Scientist Fellowship (Grant No. 2018001). J.W.Z. and X.X.Z. acknowledge the financial support from the King Abdullah University of Science and Technology (KAUST), Office of Sponsored Research (OSR) under the Award No. OSR-2017-CRG6-3427.

**Author contributions**

G.Q.Y. conceived the project. X.X. grew and characterized the bulk FGT crystal. J. T.

exfoliated the FGT thin films and fabricated the devices with the help from X. W.. X.W. performed the electrical measurements. J.Z. performed the TEM measurements. G.Q.Y. drafted the paper and all authors commented on the manuscript. The study was performed under the supervision of G.Q.Y. and X.H.. X.W., J.T., and X.X. contributed equally to this research.

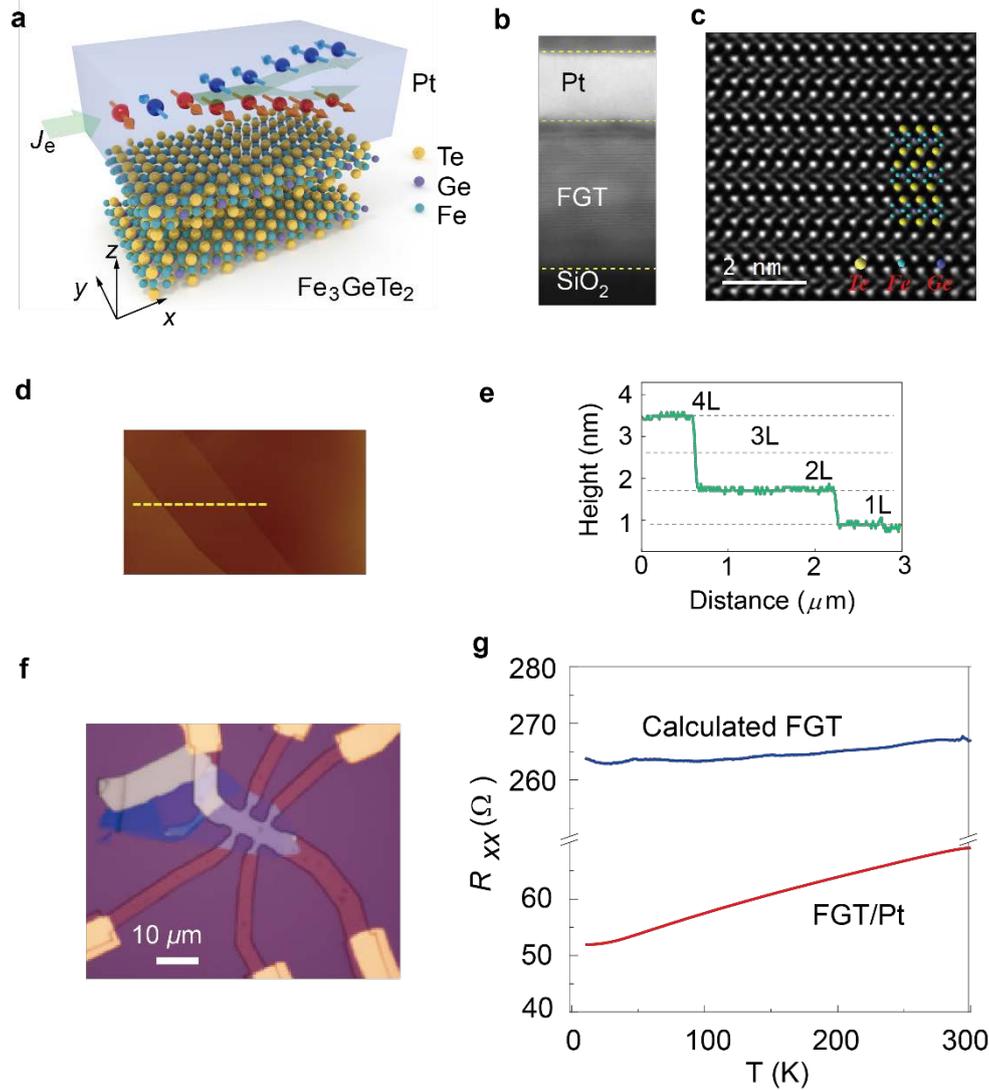

Figure 1. **Schematic view and characterizations of FGT/Pt bilayer**. **a**, Schematic view of the bilayer structure. Pt layer (top) is sputtered on top of the exfoliated FGT (bottom). The green arrow represents the in-plane current flowing in the Pt layer, which generates a spin current flowing in $z$-direction. The accumulated spins at the bottom (top) Pt surface are indicated by the red (blue) arrows. The spin current exerts torques on the magnetization of FGT and can switch it in the presence of an in-plane magnetic field. **b**, The cross-sectional scanning transmission electron microscopy (STEM) image of the FGT/Pt device fabricated on Si/SiO$_2$ substrate. The total thickness of FGT is 12.6 nm. **c**, High-resolution STEM image of a FGT(87 nm)/Pt(6

nm) bilayer on a Si/SiO$_2$ substrate. **d**, Top view of the FGT exfoliated from the bulk material measured by atomic force microscopy (AFM). **e**, The atomic steps profile taken along the yellow dashed lines in (**d**). Atomic layer step of 0.8 nm is observed. **f**. The optical image of the measured Hall bar device. **g**. Temperature-dependent longitudinal resistance of the FGT/Pt bilayer device and FGT only.

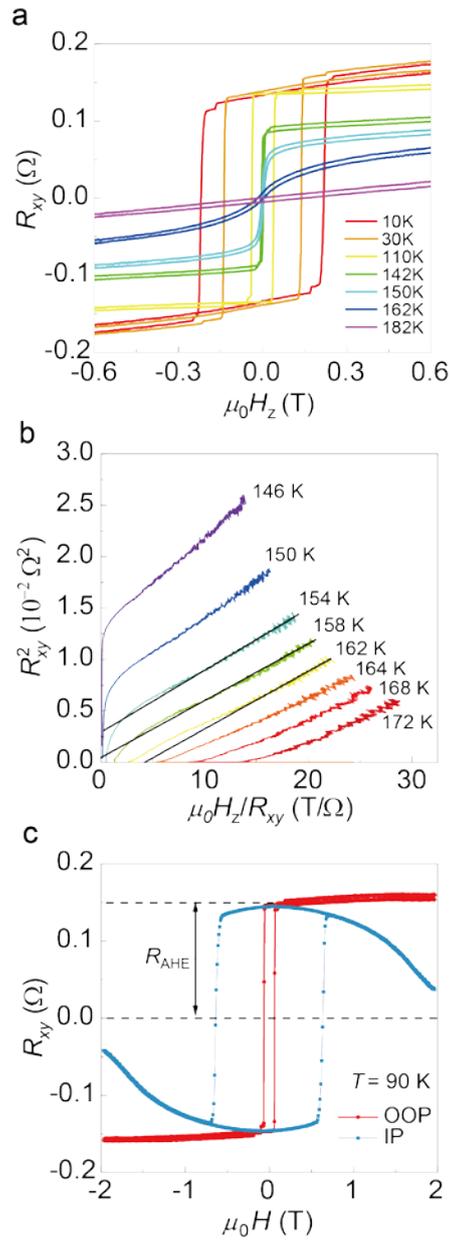

Figure 2. **Magnetic properties of FGT/Pt bilayer**. **a**, Hall resistance as a function of magnetic field at different temperatures. **b**, Arrott plots of the Hall resistance of the FGT/Pt device. The determined $T_c$ is 158 K. **c**, $R_{AHE}$ as a function of in-plane (IP) and out-of-plane (OOP) magnetic field at 90 K.

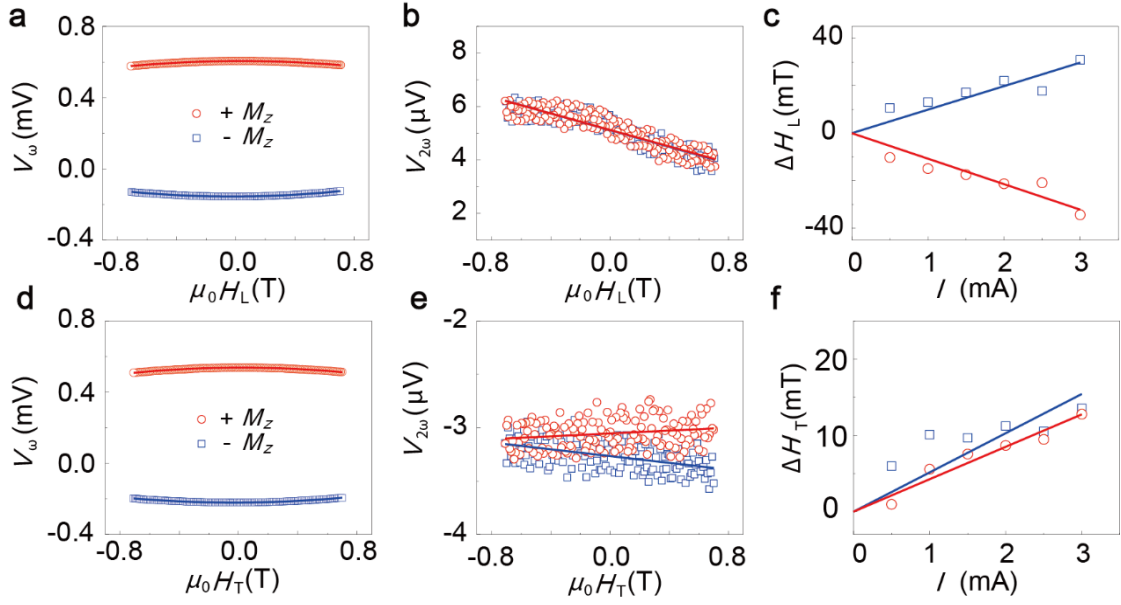

Figure 3. **Characterization of the current-induced effective fields**. **a,b**, First and second harmonic voltages for the longitudinal effective field. $H_L$ is the applied longitudinal magnetic field along the current direction ($x$ axis). **d,e**, First and second harmonic voltages for the transverse effective field. $H_T$ is the applied transverse magnetic field transverse to the current direction ($y$ axis). **c,f**, Plots of the longitudinal and transverse field as a function of the peak current. The solid lines represent the linear fitting result with zero intercept. The red circles (blue squares) are data points for the $M_z > 0$ ($M_z < 0$).

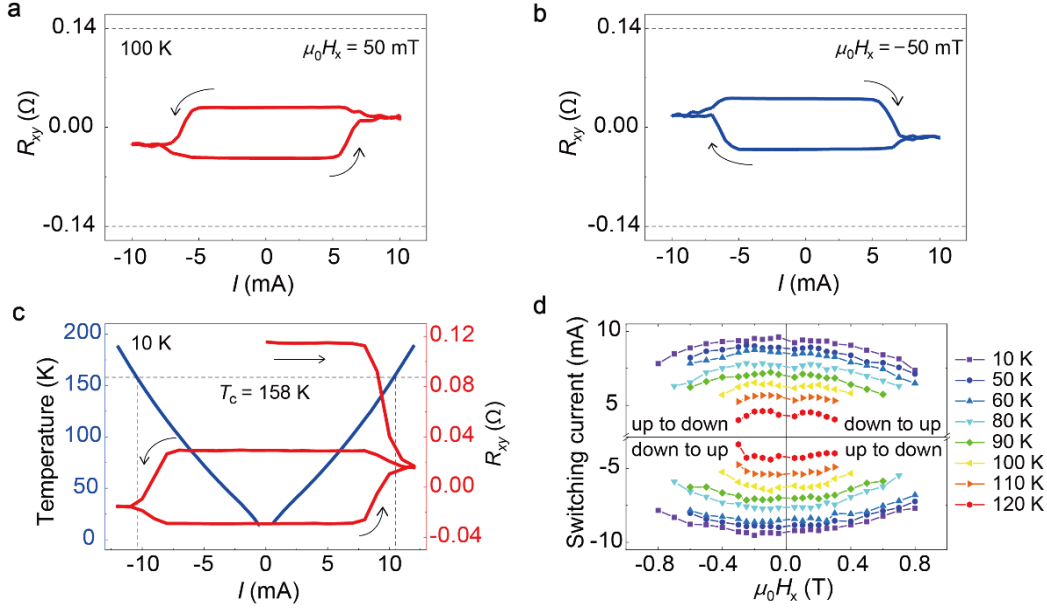

Figure 4. **Spin-orbit-torque-driven perpendicular magnetization switching in a FGT/Pt bilayer device**. Current-driven perpendicular magnetization switching for in-plane magnetic fields of 50 mT (**a**) and −50 mT (**b**) at 100 K. The switching polarity is anticlockwise and clockwise, respectively. The dashed lines correspond to the $R_{AHE}$ at saturated magnetization states. **c**, Current-driven perpendicular magnetization switching with a 300 mT in-plane magnetic field at 10 K (red). The arrows indicate the current sweeping direction. The initial state is saturated in the positive direction. The current increases gradually in the positive direction and the $R_{AHE}$ jumps down to an intermediate state. The two states in the switching loop do not correspond to the saturated states. The device temperature (blue) is obtained by comparing the measured longitudinal resistance and the measured $R_{xx}T$ curve (Fig. S14). The dashed line corresponds to the $T_c$ obtained from the Arrott plots. **d**, Switching phase diagram respect to the in-plane magnetic fields and critical switching currents at different temperatures. The critical switching current decreases with increasing temperature.